# Intrinsic synchronization of an array of spin-torque oscillators driven by the spin-Hall effect

G. Siracusano[1], R. Tomasello[2], V. Puliafito[1], A. Giordano[1], B. Azzerboni[1], A. La Corte[4], M. Carpentieri[3] and G. Finocchio[1]

[1] Department of Electronic Engineering, Industrial Chemistry and Engineering, University of Messina, C.da di Dio, I-98166, Messina, Italy

[2] Department of Computer Science, Modelling, Electronics and System Science, University of Calabria, Via P. Bucci, I-87036, Rende (CS), Italy

[3] Department of Electrical and Information Engineering, Politecnico of Bari, via E. Orabona 4, I-70125 Bari, Italy

[4] Department of Informatic Engineering and Telecommunications, University of Catania, Viale Andrea Doria 6, 95125 Catania, Italy

**Abstract**

This paper micromagnetically studies the magnetization dynamics driven by the spin-Hall effect in a Platinum/Permalloy bi-layer. For a certain field and current range, the excitation of a uniform mode, characterized by a power with a spatial distribution in the whole ferromagnetic cross section, is observed. We suggest to use the ferromagnet of the bi-layer as basis for the realization of an array of spin-torque oscillators (STOs): the Permalloy ferromagnet will act as shared free layer, whereas the spacers and the polarizers are built on top of it. Following this strategy, the frequency of the uniform mode will be the same for the whole device, creating an intrinsic synchronization. The synchronization of an array of parallely connected STOs will allow to increase the output power, as necessary for technological applications.

Spin-Torque Oscillators (STOs) are the smallest self-oscillators existing in nature.[1] Conventional STOs are made by two ferromagnets. One is designed to act as current polarizer, whereas the other one (free layer) can modify its magnetic state. They are separated by a non-magnetic material (a metal in spin-valves,[2,3] an insulator in magnetic tunnel junctions[4,5]). Besides their nano-dimensions, STOs have turned out competitive for their easy frequency tunability and narrow linewidth.[1] Moreover, synchronization of more STOs, by means of phase locking, has led to an increase of the output power and a reduction of the linewidth.[6,7,8] Since the initial demonstration, the advances in the STOs synchronization are going ahead slowly, considering that only synchronization of three high-frequency nano-contacts STOs or of four vortex oscillators has been measured so far.[9,10]

Differently from the spin-transfer torque (STT) due to a perpendicular spin-polarized current,[11] a further way to manipulate the ferromagnet magnetization (switching,[12,13] domain wall motion[14] and persistent magnetization oscillations[15,16]) has been recently observed in heavy metal/ferromagnet/insulator multilayered structures. Such dynamics are driven by the spin-orbit torque (SOT), whose origin is related to at least two key mechanisms: Rashba and Spin-Hall effects (SHE).[17,18,19,20] The Rashba term is proportional to the current flowing through the ferromagnet and to the Rashba coefficient,[21] whereas the SHE term is proportional to the current flowing through the heavy metal and to the spin-Hall angle.[14] This category of devices is very interesting from a technological point of view, being the practical realization of those structures simpler than the fabrication procedures for standard STOs.

Here, we have performed micromagnetic simulations of a Platinum (Pt) \ Permalloy (Py≡$Ni_{80}Fe_{20}$) bi-layer, where the magnetization self-oscillations are driven by the SHE. The main result of this paper is the excitation of a uniform mode, spatially distributed in the whole ferromagnetic layer, for a certain range of current and external field. This outcome is in qualitative agreement with recent experimental evidences[22] and opens perspectives for the construction of an array of intrinsically synchronized conventional STOs, which share the ferromagnet of the bi-layer as free layer. This work has been completed by the study of the effect of the Oersted field on the magnetization dynamics and its characterization using a joint time-frequency analysis.[23, 24, 25]

Our investigation has been carried out by means of a "state of the art" self-implemented micromagnetic solver,[26] which numerically integrates the Landau-Lifshitz-Gilbert-Slonczweski (LLGS) equation:

$$\frac{d\mathbf{m}}{\gamma_0 M_S dt} = -\frac{1}{(1+\alpha^2)}\mathbf{m}\times\mathbf{h}_{\mathbf{EFF}} - \frac{\alpha}{(1+\alpha^2)}\mathbf{m}\times\mathbf{m}\times\mathbf{h}_{\mathbf{EFF}}$$
$$-\frac{d_J}{(1+\alpha^2)\gamma_0 M_S}\mathbf{m}\times\mathbf{m}\times\mathbf{\sigma} + \frac{\alpha d_J}{(1+\alpha^2)\gamma_0 M_S}\mathbf{m}\times\mathbf{\sigma}$$

(1)

being $\mathbf{m}$ and $\mathbf{h}_{\mathbf{EFF}}$ the magnetization of the ferromagnet and the effective field, respectively. $\gamma_0$ is the gyromagnetic ratio, $M_s$ is the Py saturation magnetization and $\alpha$ is the Gilbert damping. The coefficient $d_J$ is given by $d_J = \frac{\mu_B \alpha_H}{eM_S t_{FE}} j_{HM}$, in which $\mu_B$ is the Bohr magneton, $e$ is the electron charge, $t_{FE}$ is the Py thickness, $\alpha_H$ is the spin-Hall angle obtained by the ratio between the amplitude of the spin current and the electrical current density $j_{HM}$ flowing through the heavy metal.[27, 28] $\mathbf{\sigma}$ is the direction of the spin-polarization in the Pt underlayer. The micromagnetic model includes, as well as the standard micromagnetic fields, the Oersted field from the current $j_{HM}$.[29, 30] The SOT due to the SHE gives rise to a Slonczewski-like torque term, which acts as a negative damping by compensating the natural losses due to the Gilbert damping, as it is possible to see in Eq. (1). The magnetic parameters used for the micromagnetic study are: exchange constant $A = 2.0 \times 10^{-11}$ J/m, $M_s = 650 \times 10^3$ A/m,[22, 31] $a = 0.02$ and $\alpha_H = 0.08$.[13]

Fig. 1 illustrates the studied device whose in-plane dimensions are $l \times w = 2000 \times 200$ nm$^2$ and the thicknesses of Py, $t_{FE}$, and Pt, $t_{HM}$, are equal to 5 nm. A cartesian coordinate system is introduced as depicted in the figure. The current $j_{HM}$ flows in the $x$-direction, whereas the external field $H_{ext}$ is applied along the $y$-direction.

The computation of the spatial distribution of the current density shows that the current flows mostly in the Pt layer (we have considered conductivities of $5.1 \times 10^6$ $(\Omega m)^{-1}$ and $2.2 \times 10^6$ $(\Omega m)^{-1}$ for Pt and Py, respectively),[22, 32] which makes the Rashba effect negligible.

Fig. 2 shows the Fourier transform of the $x$-component of the self-oscillating magnetization, when $H_{ext}$=80 mT (Fig. 2a) and 90 mT (Fig. 2b), respectively, whereas $j_{HM}$ is swept from $-0.90 \times 10^8$ A/cm$^2$ to $-1.20 \times 10^8$ A/cm$^2$. For a magnitude of $|j_{HM}| > 0.90 \times 10^8$ A/cm$^2$, no persistent precession is excited. Results for $H_{ext}$=80 mT indicate the excitation of a single mode, whose frequency increases with higher values of $|j_{HM}|$. In the case of $H_{ext}$=90 mT, for $j_{HM} = -0.90 \times 10^8$ A/cm$^2$ only an excited mode is observed with a frequency peak P1, whereas, for $j_{HM} = -1.00 \times 10^8$ A/cm$^2$, two close frequency peaks P2 and P3 arise, revealing the existence of two independent modes. No relevant self-oscillations are observed for larger magnitudes $|j_{HM}| > 1.00 \times 10^8$ A/cm$^2$.

In Fig. 3 the Fourier transform as a function of $H_{ext}$ is represented for two values of the current: $j_{HM} = -0.90 \times 10^8$ A/cm$^2$ (Fig. 3a) and $-1.00 \times 10^8$ A/cm$^2$ (Fig. 3b), respectively. For both current density values, the frequency of the self-oscillation increases as the $H_{ext}$ increases. In particular, we obtain the generation of a single mode for all the $H_{ext}$ values, except for the configuration $H_{ext}$=90 mT and $j_{HM} = -1.00 \times 10^8$ A/cm$^2$, where two modes are observed (same configuration previously represented in Fig. 2b).

In order to achieve a deeper understanding on the dynamics described above, we have studied the action of the Oersted field, the spatial mode distributions[33] (SMDs) and the results of a time-frequency analysis as performed by computing the Micromagnetic Wavelet Scalogram (MWS) according to the method described in Ref. 23.

When the Oersted field is neglected, no significant changes in the frequency spectra are detected (not shown). This is an expected result, since in such kind of bi-layered device the Oersted field is uniform in the whole ferromagnet and it is oriented in the same direction of $H_{ext}$.

On the contrary, in conventional STOs, the Oersted field is non-uniform, exhibiting a higher amplitude near the contacts and lower towards the boundaries. When the lateral dimensions of STOs are increased, it will exist a critical size over which the Oersted field will become predominant on the STT, leading to the nucleation of non-uniform magnetization patterns, such as vortexes.

Fig. 4 reports the time-frequency study based on the Wavelet Analysis (WA), as computed by means of MWS (the amplitude of the wavelet coefficient increases from white to black), whereas the SMDs are calculated using micromagnetic spectral mode decomposition mapping technique (MSMDMT).[34] In detail, Fig. 4a represents the results for the case $H_{ext}$=90 mT and $j_{HM} = -0.90 \times 10^8$ A/cm$^2$. In this scenario, when a single mode is observed (inset of Fig. 4a), it is uniform in the whole ferromagnetic cross section (its frequency is $f_{P1} = 6.2$ GHz), and WA reveals that the mode P1 is also stationary in time. Fig. 4b refers to the configuration $H_{ext}$=90 mT and $j_{HM} = -1.00 \times 10^8$ A/cm$^2$. As it can be clearly seen, the two modes P2 ($f_{P2}$=6.4 GHz) and P3 ($f_{P3}$=6.5 GHz) are uniform (see insets in Fig. 4b) but non-stationary in time, being excited in two different time windows. In addition, no differences in both the SMDs and WA are found when the Oersted field is neglected, as well as in the computation of the power which highlights a non-linear trend with current (not shown).[7]

In summary, our results are interesting from a technological point of view, since when a uniform mode is generated it is possible to design an embedded microwave oscillator made by an array of conventional STOs intrinsically synchronized. The architecture would be similar to the array of three-terminal MTJs,[16,35] where, likewise, the self-oscillations were achieved through the SHE and the output power was read via the tunneling magnetoresistance. Differently from that geometry, the device here proposed is an array of STOs built over the heavy metal and sharing also the free layer (Py ferromagnet), with the advantage to attain the synchronization without applying any external microwave source.

**Acknowledgements**

This work was supported by the Italian MIUR under Grant PRIN2010ECA8P3. The work of R. Tomasello and M. Carpentieri was also supported by the Fondazione Carical through the Project entitled "Progettazione e Sviluppo di Sensori Spintronici per Imaging a Microonde con Applicazione ai test Non Distruttivi e alla Caratterizzazione dei Materiali".

**Figures captions**

**Fig. 1**: Schematic representation of the heavy metal\ferromagnet bi-layered device. (Inset) Detailed sketch of the by-layer structure showing the thicknesses of the layers, the direction of the current density $j_{HM}$ and the applied external magnetic field $H_{ext}$.

**Fig. 2**: Fourier transform as a function of $j_{HM}$. (a) $H_{ext}$=80 mT. (b) $H_{ext}$=90 mT.

**Fig. 3**: Fourier transform as a function of $H_{ext}$. (a) $j_{HM} = -0.90 \times 10^8$ A/cm$^2$, (b) $j_{HM} = -1.00 \times 10^8$ A/cm$^2$.

**Fig. 4**: MWS for $H_{ext}$=90mT, where the power increases from white to black. (a) $j_{HM} = -0.90 \times 10^8$ A/cm$^2$, (b) $j_{HM} = -1.00 \times 10^8$ A/cm$^2$. Insets: SMDs related to the *x*-component of the magnetization (the power increases from white to red).

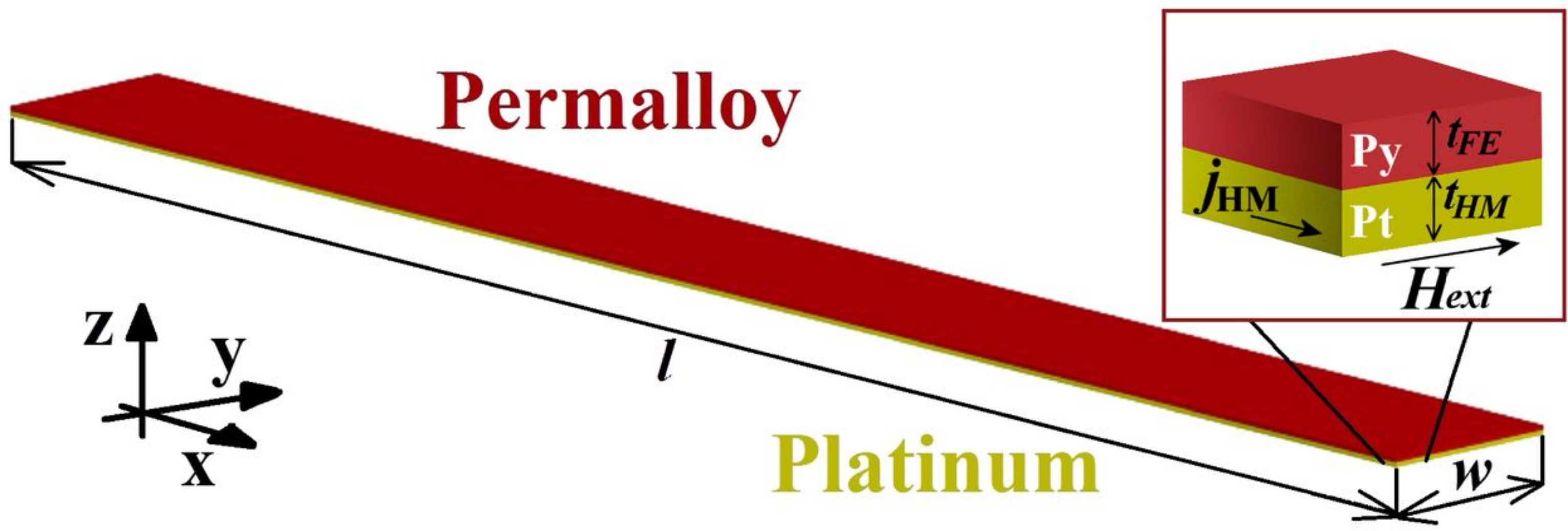

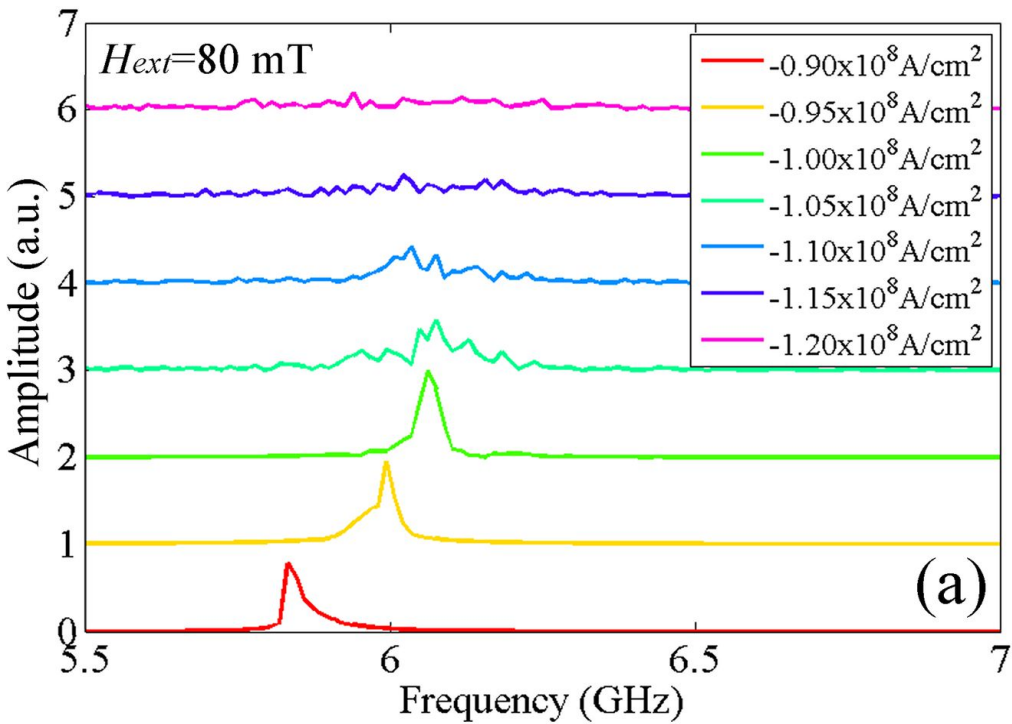
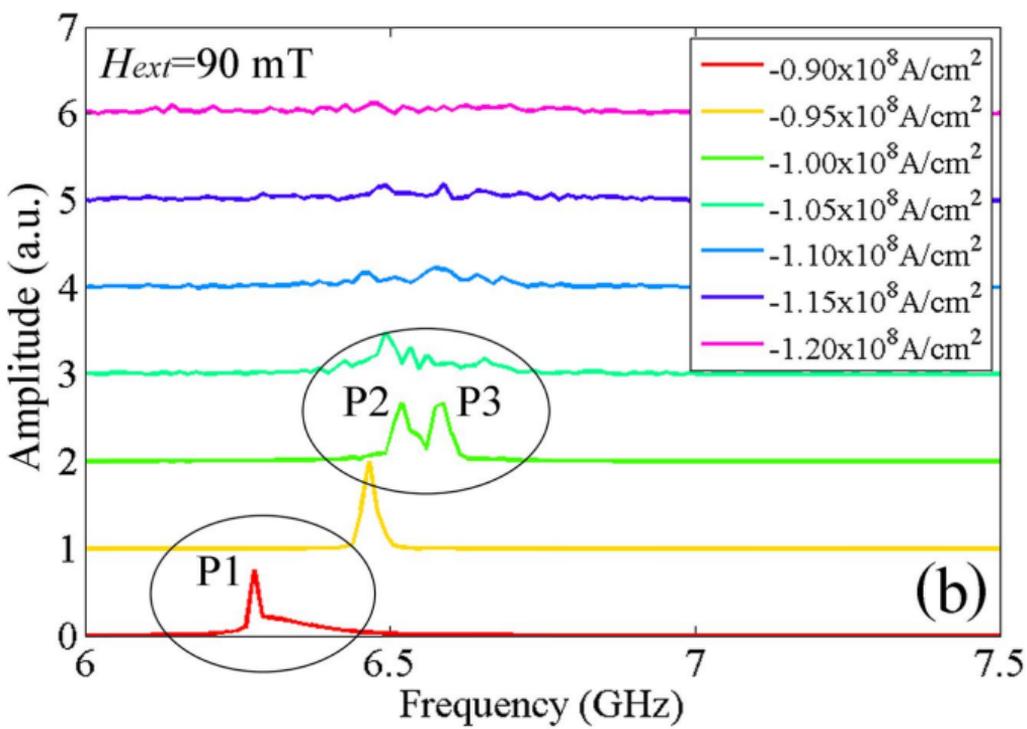

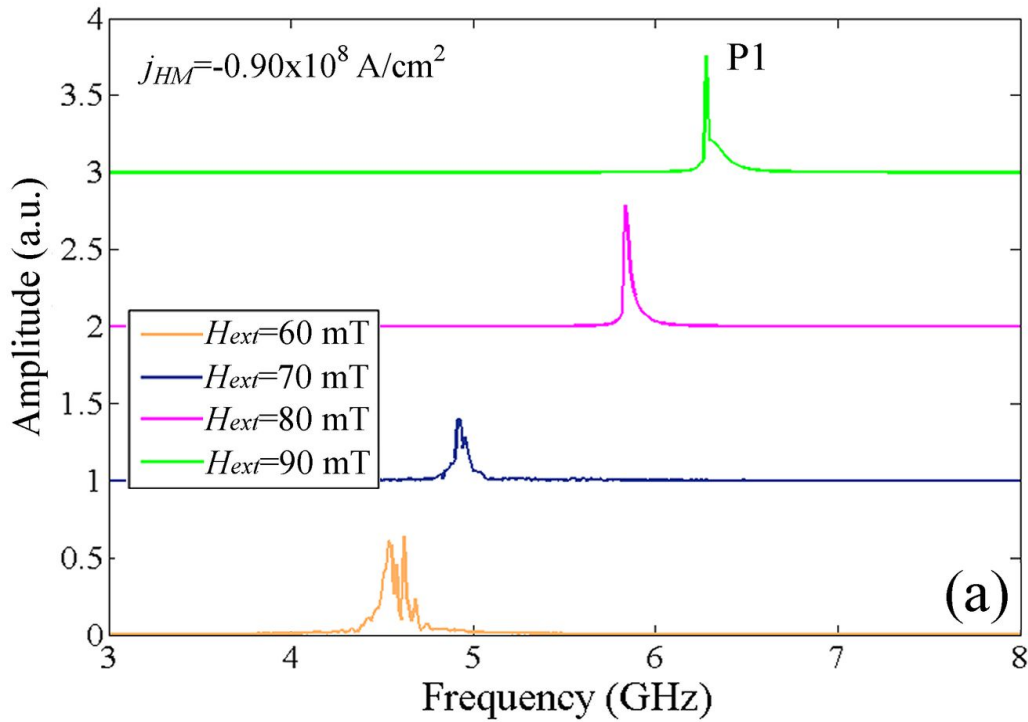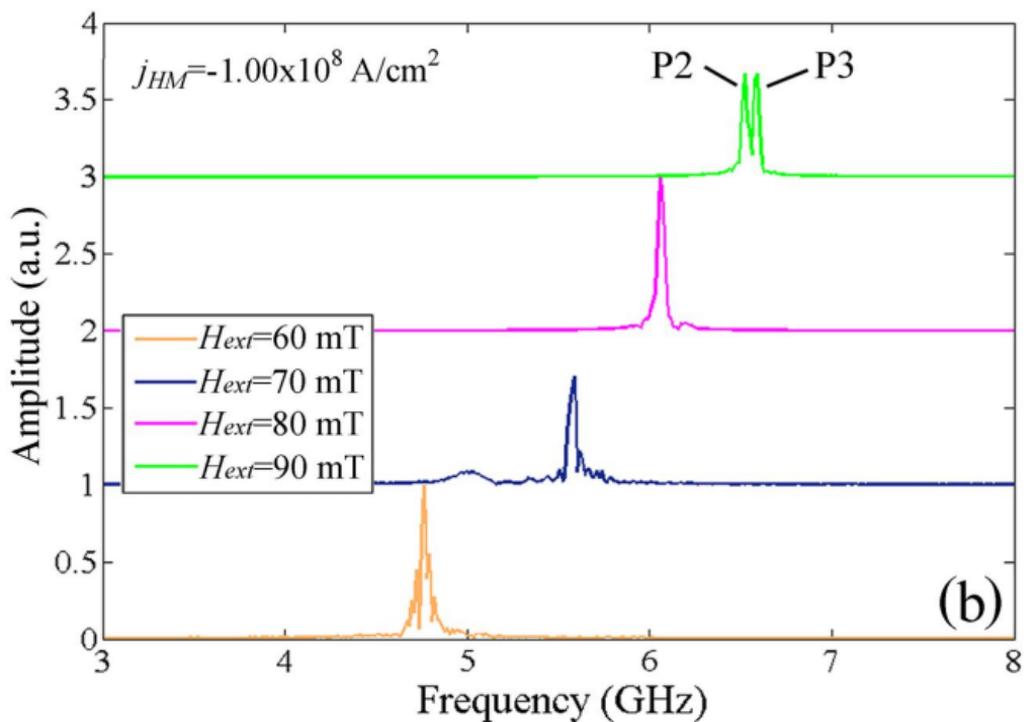

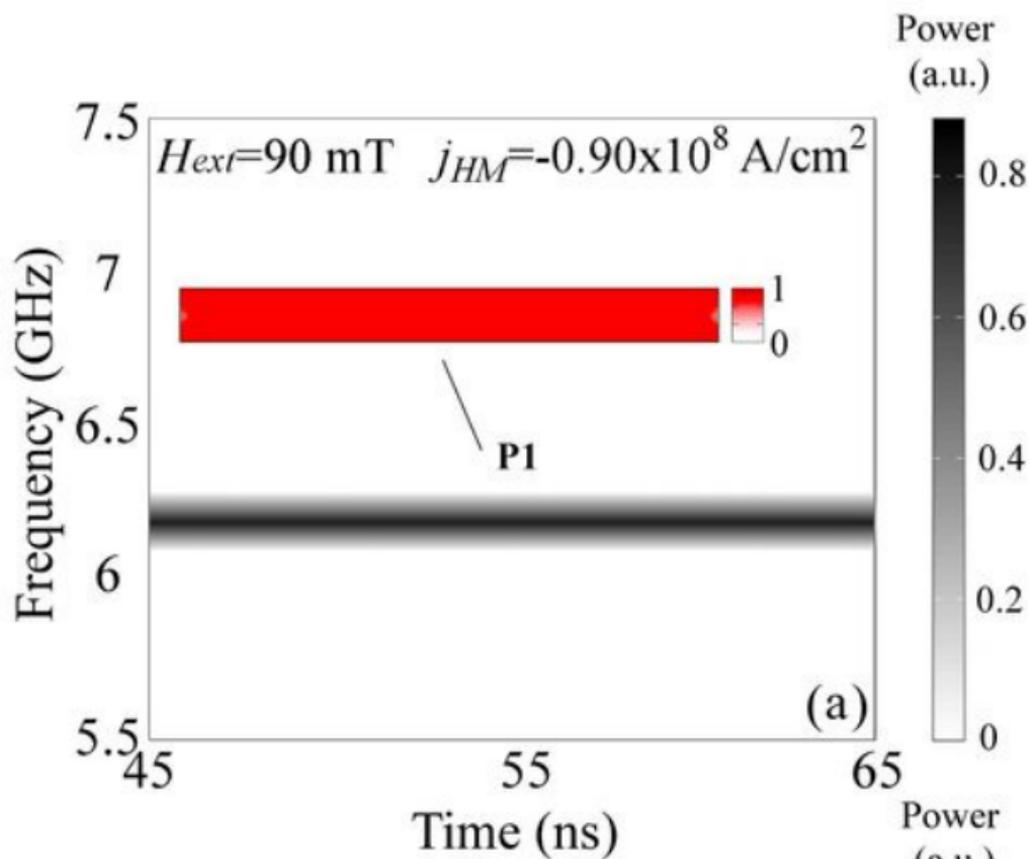
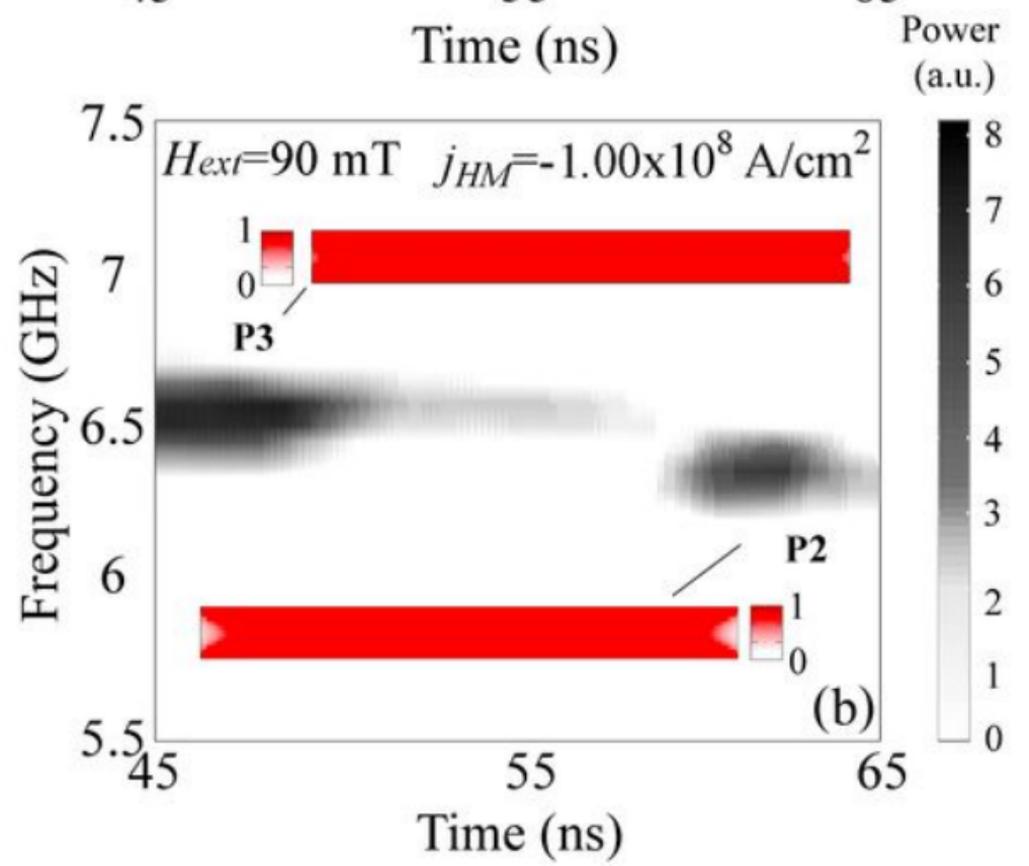